# Lower Hybrid Antennas for Nuclear Fusion Experiments

J. Hillairet, J. Achard, Y.S. Bae[1], X. Bai[2], C. Balorin, Y. Baranov[3], V. Basiuk, A. Bécoulet, J. Belo[4], G. Berger-By, S. Brémond, C. Castaldo[5], S. Ceccuzzi[5], R. Cesario[5], E. Corbel, X. Courtois, J. Decker, E. Delmas, L. Delpech, X. Ding[2], D. Douai, A. Ekedahl, C. Goletto, M. Goniche, D. Guilhem, J.P. Gunn, P. Hertout, G.T. Hoang, F. Imbeaux, K.K. Kirov[3], X. Litaudon, R. Magne, J. Mailloux[3], D. Mazon, F. Mirizzi[5], P. Mollard, P. Moreau, T. Oosako, V. Petrzilka[6], Y. Peysson, S. Poli, M. Preynas, M. Prou, F. Saint-Laurent, F. Samaille, B. Saoutic, P.K. Sharma

CEA, IRFM, F-13108 Saint Paul-lez-Durance, France
julien.hillairet@cea.fr
1 National Fusion Research Institute, Daejeon, South Korea
2 Southwestern Institute of Physics, Chengdu, People's Republic of China
3 Euratom/CCFE Fusion Association, Culham Science Centre, Abingdon, OX14 3DB, UK
4 Associacao Euratom-IST, Centro de Fusao Nuclear 1049-001 Lisboa, Portugal
5 Associazione Euratom-ENEA sulla Fusione, CR Frascati, Roma, Italy
6 Association Euratom-IPP.CR, Za Slovankou 3, 182 21 Praha 8, Czech Republic

*Abstract*— The nuclear fusion research goal is to demonstrate the feasibility of fusion power for peaceful purposes. In order to achieve the conditions similar to those expected in an electricity-generating fusion power plant, plasmas with a temperature of several hundreds of millions of degrees must be generated and sustained for long periods. For this purpose, RF antennas delivering multi-megawatts of power to magnetized confined plasma are commonly used in experimental tokamaks. In the gigahertz range of frequencies, high power phased arrays known as "Lower Hybrid" (LH) antennas are used to extend the plasma duration. This paper reviews some of the technological aspects of the LH antennas used in the Tore Supra tokamak and presents the current design of a proposed 20 MW LH system for the international experiment ITER.

*Nuclear Fusion; high power; phased array; rectangular waveguide; Lower Hybrid; Current Drive; Tore Supra; ITER*

## I. Introduction

The ultimate goal of nuclear fusion research is to demonstrate the scientific and the technological feasibility of fusion power for peaceful purposes. In order to achieve the conditions similar to those expected in an electricity-generating fusion power plant, very hot plasmas, with temperature exceeding 100 million of degrees, must be generated and sustained for long periods.

In order to increase the plasma temperature and to achieve long pulse operations, a *tokamak* -- such as the international experimental project ITER which is currently under construction in Cadarache (France) -- requires additional heating and current drive systems. Radio Frequency (RF) antennas, delivering multi-megawatts level of power into the plasma in which their energy is transferred to the charged particles, are presently used in many different *tokamaks* in the world. The magnetised plasma is an inhomogeneous, anisotropic and lossy medium in which many wave modes can co-exist depending on their frequency and their polarization. RF heating can be achieved by launching a wave at the ion or electron cyclotron frequency. For *tokamaks*, ion heating requires a frequency in the tens of MHz range, electron heating in the 40-200 GHz range. The GHz range of frequencies which is in between is used by the so-called "Lower Hybrid" (LH) systems. LH antennas are phased arrays of open-ended rectangular waveguides which are phased to launch traveling waves into the plasma. The launched LH waves accelerate resonant electrons and drive a current inside the plasma which helps to maintain the plasma discharge for long periods.

In the superconducting experimental *tokamak* Tore Supra (Cadarache, France), 7 MW CW generated by sixteen 3.7 GHz klystrons are transferred to two antennas which are facing the plasma edge [1]. This allowed Tore Supra to set up a record in 2003 with a plasma duration of more than six minutes during which energy on the order of 300 kWh was injected and extracted [2]. In the next-generation *tokamak* ITER, a 20 MW Lower Hybrid Range of Frequency launcher at 5 GHz aiming to extend the plasma performance and duration is proposed, and a pre-design using the RF schemes of present LH antennas has been performed [3].

In this paper we describe the LH system installed on Tore Supra from its RF sources (klystrons), transmission-lines (rectangular waveguides) and antennas. The latest results obtained with an ITER-relevant LH antenna are exposed and the preliminary design of the proposed ITER LH antenna is also outlined.

## II. LOWER HYBRID CURRENT DRIVE IN TOKAMAKS

In a *tokamak*, the magnetic configuration which confines the plasma results from the addition of a toroidal and a poloidal magnetic field. The toroidal field is generated by coils surrounding the plasma chamber while the poloidal field comes from a large DC current (up to 15 MA in ITER) flowing into the plasma ring. This DC current is inductively generated by means of a transformer, which is intrinsically a pulsed mode process. In order to reach steady-state operations, some means of external non-inductive current drive is required. From all the methods used to generate additional plasma current, the lower hybrid current drive (LHCD) method has the highest driven current per watt of applied power (close to 0.05 A/W for ITER).

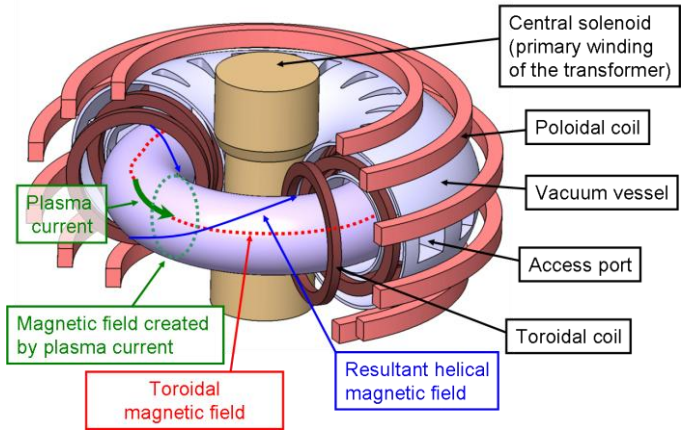

Figure 1. Illustration of the magnetic field configuration created in a *tokamak*.

The magnetic confined plasma is a medium which is generally inhomogeneous, anisotropic, lossy and dispersive. Near the plasma edge however, the lossy and dispersive properties can be neglected as a first approximation, in a so-called "cold plasma" model[4]. The dielectric tensor in such plasma is not diagonal in general, because particles act very differently depending if their dynamics are along or perpendicular to the magnetic field. This anisotropy leads to define parallel $k_{//}$ and perpendicular $k_\perp$ components of the wavevector **k** with respect to the DC magnetic field $\mathbf{B}_0$, and so to the vector refractive index $\mathbf{n}=\mathbf{k}c/\omega$. Since no variation of the DC magnetic field $\mathbf{B}_0$ or of the particle density is assumed along the parallel direction, the $k_{//}$ value is determined by the geometric structure of the antenna as the waves enter the plasma.

The main aim of LH antennas is to use a wave-particle collisionless damping mechanism acting on electrons, known as Landau damping, in order to generate an additional current into the plasma. This damping is achieved through quasi-electrostatic waves whose electric field is parallel to $\mathbf{B}_0$. Such fields are excited by means of a phased array of narrow rectangular waveguides, which broad sides are perpendicular to the magnetic field [5]. Efficient numerical tools allow predicting the power coupling efficiency of such antennas to *tokamak* edge plasmas [6]. For appropriate chosen geometry and RF parameters such as relative phase between waveguides, waves propagate predominantly in one direction along the torus. The electrons dragged with these waves ultimately produce some plasma current [7].

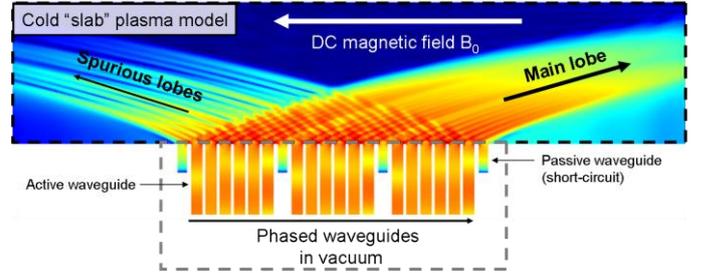

Figure 2. Illustration of the magnitude of electric field parallel to the DC magnetic field at the interface between a a Lower Hybrid antenna (cut-view) and the *tokamak* plasma edge (log-scale). This example antenna is made of 18 active waveguides and 4 passive waveguides (short-circuits). The power is propagating with a preferred direction – the opposed direction comes from the spurious lobes generated.

## III. TORE SUPRA LOWER HYBRID SYSTEM

Tore Supra is one of the largest *tokamak* in the world, in operation since 1988. Main features are its superconducting toroidal magnets and its actively water-cooled first wall allowing to study plasma performance in steady-state conditions. This experimental device located in Cadarache (France) is specialized to the study of physics and technology dedicated to long plasma discharges heated by large RF powers ($P_{RF}$>10MW).

The maximum RF power which is possible to transmit to the plasma via rectangular waveguides is partly limited by electric breakdowns into the waveguides. In order to achieve multi-megawatts range of transmitted RF power, hundreds of waveguides are used for an antenna. If each waveguide would be fed by a dedicated source, this would obviously lead to a very complex and expensive RF architecture, either from the sources or for the transmission lines viewpoint. In order to avoid such complexity, the RF power which is generated by klystrons is split by passive waveguide components such as Riblet hybrid junction[8], mode converters [9], and multijunction[9]. A multijunction is an E-plane power splitter which consists in an assembly of E-plane bi-junctions. Some phase shifters are inserted in the rectangular waveguides, by reducing locally the height of the waveguide in order to increase the phase speed of the $TE_{10}$ mode (Figure 3. ). The whole device is thus a phased array of rectangular waveguides where the phase shift between two adjacent waveguides is fixed by design. The use of multijunction antenna allows not only to simplify the antenna feeding transmission lines, but also to reduce the amount of reflected power back to the RF sources. This self-matching property is due to the fact that the reflected power from the plasma to the antenna experiences multiple destructive interferences inside the multijunction[11].

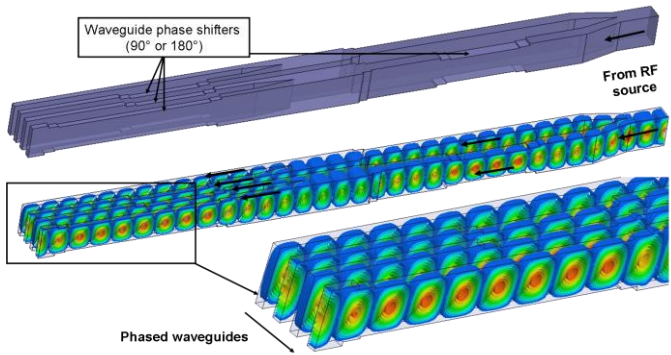

Figure 3. Illustration of a multijunction. A multijunction is an assembly of rectangular waveguides and E-plane splitters. The resulting phase shift between output waveguides at the left is mechnically fixed by the geometry of the multijunction phase shifters.

In Tore Supra, two LH antennas are installed and each of them is fed by eight 3.7 GHz 700 kW CW klystrons from Thales Electron Device. Figure 4. illustrates both antennas front face as it could be seen inside the *tokamak* vacuum chamber. The most recent antenna (C4) has been installed in 2009. Its concept differs from previous antenna by the fact that an equivalent short circuit ("passive" waveguide) is inserted between all directly RF fed waveguides ("active" waveguides) of the antenna. This scheme, called Passive-Active Multijunction (PAM) [12][13], allows a better coupling of the power especially when the plasma-wave mode is close to a cutoff regime, which occurs when the electron density near the antenna is close or lower to a specific cut-off density [13]. These performances are due to the increase of the distance between two active waveguides, which leads to better radiating properties in the vacuum or below the cut-off density. This is of particular interest for ITER and future fusion demonstrator, in which the distance between the antenna and the plasma edge would be greater than for present device, thus leading to smaller edge particle density at the mouth of the antenna. Moreover, cooling pipes are vertically drilled behind the passive waveguides in order to actively cool the antenna front face and the waveguide walls and damp part of the neutron flux, which is a mandatory requirement for any ITER plasma facing component.

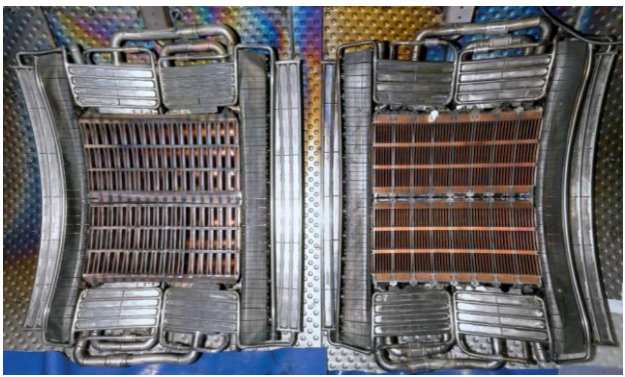

Figure 4. Pictures of the two 3.7GHz LH antennas currently in operation at Tore Supra as seen from inside the vacuum vessel. Each antenna is approximately 600x600 mm large. Left antenna has 96 active waveguides and 102 passive waveguides. Right antenna has 288 active waveguides. Each antenna is protected from unconfined high energy particles by actively water-cooled lateral bumpers, made in carbon fiber composite tiles.

## IV. PROPOSED LH ANTENNA DESIGN FOR ITER

A LH system able to deliver 20MW/CW has been proposed for the ITER *tokamak* [15]. In this design, the antenna is made of 48 identical modules, each one independently fed by one 5 GHz/500 kW CW klystron: twelve in the horizontal direction and four in the vertical direction. A module consists of four active waveguides in the horizontal direction and six ($2 \times 3$) lines of waveguides in the vertical direction (Figure 5. ). Thus, the whole antenna contains 1152 active waveguides whose dimensions are 9.25×58 mm.

The RF power is carried through a transmission line from each klystron up to a 500 kW RF window located outside the plug frame and connected to a 3 dB splitter which feeds two $TE_{10}$-$TE_{30}$ mode converters. After the mode converter, the power is divided into three vertical rows, corresponding to the inputs of a four-active waveguides PAM.

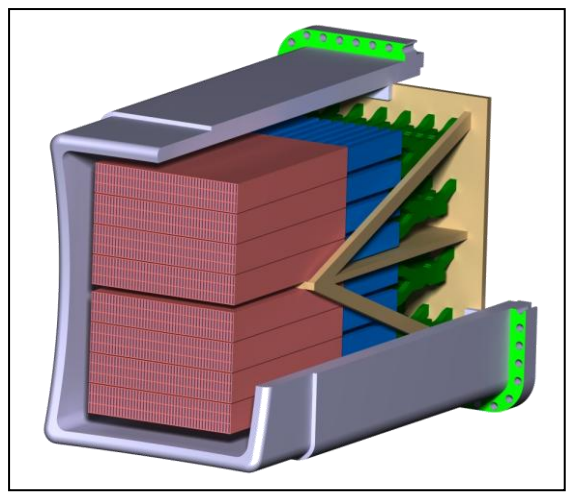

Figure 5. Illustration of the ITER LH antenna design. Antenna overall dimensions are: 2160mm height, 650mm width and 3470mm length. The red parts are the Passive-Active Multijunction, the blue parts the TE10-TE30 mode converters and the green parts the 3dB splitters. RF windows are not illustrated on this picture and are located behind the backplate illustrated in brown.

ACKNOWLEDGMENT

This work, supported by the European Communities under the contract of Association between EURATOM and CEA, was carried out within the framework of the European Fusion Development Agreement. The views and opinions expressed herein do not necessarily reflect those of the European Commission.

## REFERENCES

[1] A.Ekedahl et al., Validation of the ITER-relevant passive-active-multijunction LHCD launcher on long pulses in Tore Supra, Nuclear Fusion 50, 2010.

[2] J.Bucalossi et al., The Gigajoule Discharges, Fusion Science and Technology, Volume 56, Number 3, pp. 1366-1380, October 2009.

[3] Hoang G.T. et al, A lower hybrid current drive system for ITER, Nucl. Fusion 49 075001, 2009

[4] T.H.Stix, Waves in Plasmas, Springer, 1992.

[5] M. Brambilla, Slow-wave launching at the lower hybrid frequency using a phased waveguide array, Nucl. Fusion 16 47, 1976.

[6] J. Hillairet et al, ALOHA: an Advanced LOwer Hybrid Antenna coupling code, Nucl. Fusion 50 125010, 2010.

[7] N.J. Fisch, Theory of current drive in plasmas, Rev. Mod. Phys. 59, 175–234, 1987.

[8] H.J.Riblet, The short-slot hybrid junction, Proceedings of the I.R.E., vol. 40, pp. 180--184, Feb. 1952.

[9] Ph.Bibet et al., Experimental and theoretical results concerning the development of the main rf components for nest tore supra LHCD antennae, in Proceeding of the 18th SOFT Conference, vol. 1, 1994.

[10] Ph.Bibet et al, New advanced launcher for lower hybrid current drive on Tore Supra 5th Int. Symp.on Fusion Nuclear Technology Fusion Eng. Des.51–52 741–6, 2000.

[11] X.Litaudon et al. Coupling of slow waves near the lower hybrid frequency in JET, Nucl. Fusion 30 471, 1990.

[12] Ph.Bibet et al, Conceptual study of a reflector waveguide array for launching lower hybrid waves in reactor grade plasmas Nucl. Fusion 35 1213–23, 1995.

[13] R.W. Motley and W.M. Hooke, Active-passive waveguide array for wave excitation in plasmas, Nucl. Fusion 20 222, 1980.

[14] M.Preynas et al., Coupling characteristics of the ITER-relevant lower hybrid antenna in Tore Supra: experiments and modelling, Nucl. Fusion 51, 2011.

[15] J. Hillairet, et al., RF modeling of the ITER-relevant lower hybrid antenna, Fusion Eng. Des., 2011.